\documentclass[aps,prc,twocolumn,showpacs,amsmath,amssymb,preprintnumbers,10pt]{revtex4-1}

\usepackage[bookmarks=false]{hyperref}
\usepackage{latexsym} 
\usepackage{tikz}
\usepackage{subfigure}
\usepackage{enumerate}
\usepackage{graphicx}       
\usepackage{bbm}       
\usepackage{color}
\usepackage{bm}   

\newcommand{\De}{\Delta}

\newcommand{\eps}{\epsilon}

\renewcommand{\th}{\theta}   

\newcommand{\om}{\omega}

\newcommand{\beq}{\begin{equation}}
\newcommand{\eeq}{\end{equation}}
\newcommand{\ba}{\begin{array}}
\newcommand{\ea}{\end{array}}
\newcommand{\bea}{\begin{eqnarray}}
\newcommand{\eea}{\end{eqnarray}}
\newcommand{\bi}{\begin{itemize}}  
\newcommand{\ei}{\end{itemize}}
\newcommand{\ben}{\begin{enumerate}} 
\newcommand{\een}{\end{enumerate}}
\newcommand{\bc}{\begin{center}}
\newcommand{\ec}{\end{center}}

\newcommand{\<}{\langle}
\renewcommand{\>}{\rangle} 

\newcommand{\dsp}{\displaystyle}
\newcommand\eqn[1]{(\ref{#1})}      

\renewcommand{\O}{{\cal O}}

\newcommand{\tpt}{${}^3\!P_2$}
\newcommand{\osz}{${}^1\!S_0$}

\usepackage[normalem]{ulem}  

\begin{document}

\title{Gap-bridging enhancement of modified Urca processes in nuclear matter}

\author{Mark G. Alford and  Kamal Pangeni}

\affiliation{Physics Department, Washington University, St.~Louis, MO~63130, USA}

\begin{abstract}
In nuclear matter at neutron-star densities and temperatures,
Cooper pairing leads to the formation of a gap in 
the nucleon excitation spectra resulting in
exponentially strong Boltzmann suppression of many transport coefficients. 
Previous calculations have shown evidence 
that density oscillations of sufficiently
large amplitude can overcome this suppression
for flavor-changing $\beta$ processes, via the mechanism
of ``gap bridging''. We address the simplifications made
in that initial work, and show that gap bridging can counteract
Boltzmann suppression of neutrino emissivity 
for the realistic case of
modified Urca processes in matter with \tpt\ neutron pairing.
\end{abstract}

\date{16 Nov 2016} 

\maketitle

\section{Introduction}
\label{sec:intro}

Ultra-dense nuclear matter is believed to be a superfluid
(via neutron Cooper pairing) and a superconductor (via proton
Cooper pairing) for at least part of
the range of densities that is relevant for
neutron star physics \cite{dean2003pairing, Dickhoff2005, Ding2016}.
This has a profound effect on transport properties, many of which
are suppressed as $\exp(-\De/T)$
by the gap $\De$ in the neutron or proton spectrum.
Since neutron star core temperatures $T$ are of order
$0.01$\,MeV~\cite{Lattimer536} and $\De$ is typically in the MeV range~\cite{dean2003pairing}, 
the suppression factor can be as strong as $10^{-40}$.

It has previously been shown \cite{Alford:2011df} that 
compression oscillations of sufficiently high amplitude
can entirely overcome this suppression
for certain transport properties, such as bulk
viscosity and neutrino emissivity, that are dominated by
flavor-changing $\beta$ (weak interaction) processes.
The mechanism, called ``gap bridging'',
is a threshold-like behavior, separate from 
high-amplitude suprathermal enhancement of
$\beta$ processes
\cite{Madsen:1992sx,Reisenegger:1994be,Alford:2010gw}.
Additional enhancement may come from the suppression of the gap itself by
high-velocity flow of the superfluid relative to the normal fluid
\cite{Gusakov:2012eb}.

The previous calculation of gap bridging \cite{Alford:2011df}
found that oscillations with density amplitude that reached
$\De n/\bar n \sim 10^{-4}$ could show gap bridging enhancement.
Gap bridging is therefore expected to be relevant to high-amplitude oscillations
of neutron stars such as f- or r-modes \cite{Stergioulas:2003yp,alford2014gravitational},
or the oscillations caused by star quakes \cite{franco2000quaking} or 
neutron star mergers \cite{tsang2012shattering}. 
However, the previous calculation was an illustrative proof of principle
in which the neutron pairing was assumed isotropic, in the
\osz\ channel, and only direct Urca processes were considered.
In this paper we provide a more realistic calculation, obtaining
the gap-bridging enhancement of the modified Urca
neutrino emissivity for nuclear matter with \tpt\ neutron pairing.
We find that gap bridging is just as dramatic in this realistic
scenario as the original estimates indicated.

In Sec.~\ref{sec:mUrca} we describe the modified Urca process and the 
quantities that we will calculate to characterize the neutrino emissivity
of nuclear matter. In Sec.~\ref{sec:1S0} we calculate the 
modified-Urca emissivity
for matter with \osz\ pairing of neutrons.
In Sec.~\ref{sec:3P2} we calculate the modified-Urca emissivity
for matter with \tpt\ pairing of neutrons. Sec.~\ref{sec:conclusion} contains
our conclusions.

\section{Modified Urca process}
\label{sec:mUrca}

Urca processes change the flavor (isospin) of nucleons
and emit neutrinos. They dominate certain physical properties
such as bulk viscosity and neutrino emissivity.
In this paper we will calculate the enhancement
of the neutrino emissivity by gap bridging in high amplitude
compression oscillations.

\begin{figure}[hb]
\bc
\includegraphics[width=0.6\hsize]{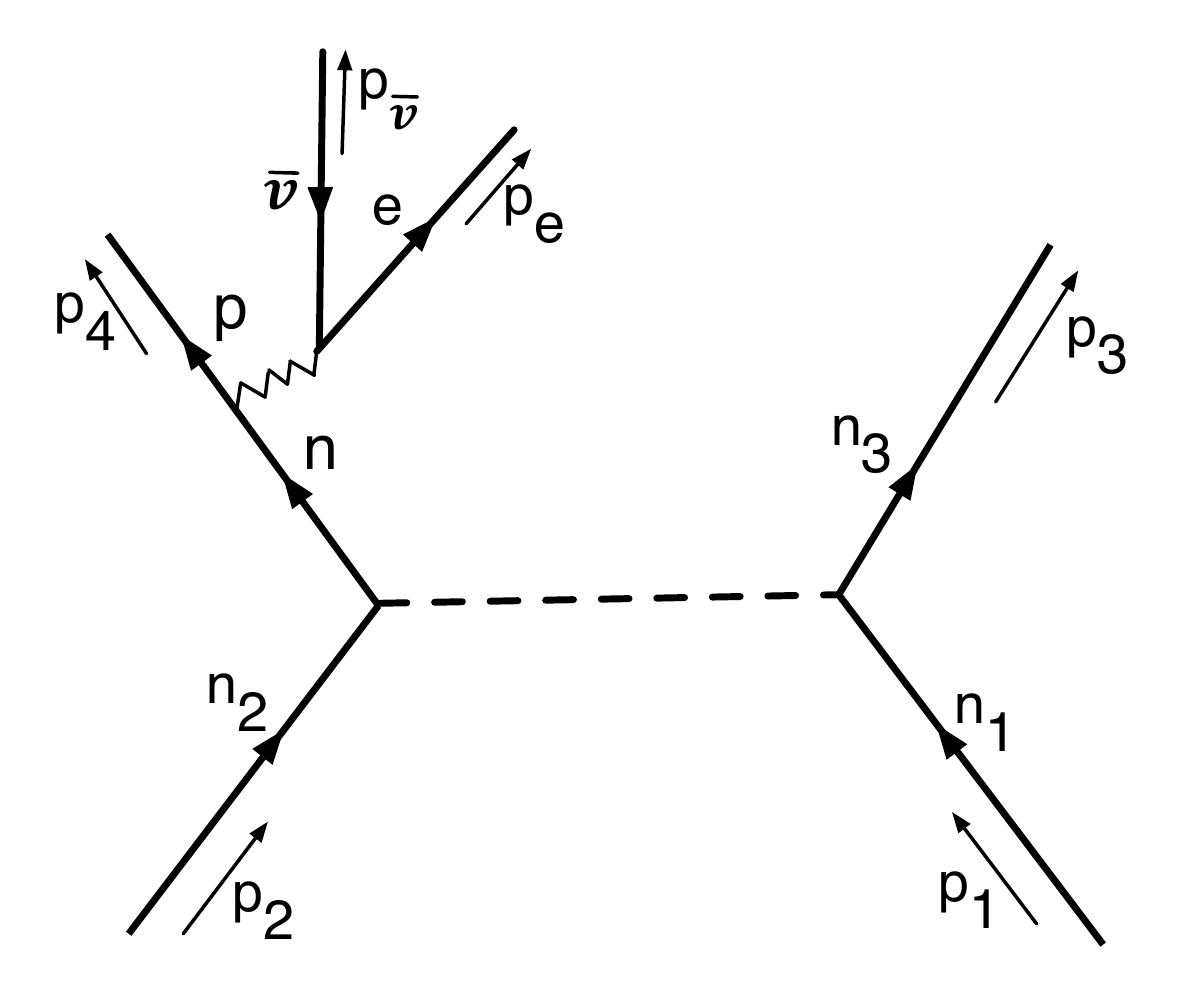}
\ec
\caption{Feynman diagram for a modified Urca process.
The initial state contains two neutrons: $n_2$, which undergoes $\beta$ decay
to a proton, and $n_1$, which is a spectator, interacting with
the other neutron or proton via the strong nuclear force.
}
\label{fig:Feynman}
\end{figure}

Initial work on gap bridging studied the direct Urca process because of
its simplicity, but in $\beta$-equilibrated nuclear matter direct Urca
processes occur only when the density reaches several
times the nuclear saturation density, at which point the proton and neutron
Fermi momenta are sufficiently similar to allow direct conversion
of one species into the other.
In most if not all regions of a neutron star
the direct Urca process $n\to p\ e^-\ \bar\nu_e$
is forbidden by energy and momentum conservation:
a neutron near its Fermi momentum $p_{Fn}$ cannot turn into
a proton near its Fermi momentum $p_{Fp}$  and an electron near its Fermi
momentum $p_{Fe}$ because $p_{Fn}>p_{Fp}+p_{Fe}$. 
In the absence of the direct Urca process, the main flavor-changing
$\beta$ process is the modified Urca process in which a
``spectator'' neutron, interacting via pion exchange, absorbs
the extra momentum (Fig.~\ref{fig:Feynman})
\begin{equation}\label{eq:murca}
\begin{split}
  n+n &\rightarrow n+p+e^{-}+\bar{\nu}_e \\
  p+n+e^{-} &\rightarrow n+n+ \nu_e
\end{split}
\end{equation}
We neglect the modified Urca process that uses a spectator proton
because it is suppressed by the lower density of protons. 

The neutrino emissivity (energy radiation rate per unit volume)
arising from the modified Urca process \eqn{eq:murca} is \cite{yakovlev2001neutrino}
\begin{equation}
\begin{split}
\label{eq:emm}
 \eps= &\int\left[ \prod_{j=1}^{4}\frac{d^3P_j}{(2\pi)^3}\right]
  \frac{d^3P_e}{(2\pi)^3}\frac{d^3P_\nu}{(2\pi)^3} (2\pi)^4
  \delta(E_{\rm f}-E_{\rm i})\times \\
& \times \delta^3{(\vec{P_{\rm f}}-\vec{P_{\rm i}})}E_\nu f_1f_2(1-f_3)(1-f_4)(1-f_e) |M_{\rm f\,i}|^2
\end{split}
\end{equation}
where the index $j$ labels the four nucleon states (two neutrons in the initial
state ``${\rm i}$'' and a proton and a neutron in the final state ``${\rm f}$'');
$f_j\equiv 1/\{1+\exp[(E_j-\mu_j)/T]\}$ are the Fermi-Dirac occupation
distributions for the nucleons, $P_j$ are the nucleon momenta, 
$P_e$ and $E_e$ are the electron momentum and energy, 
$P_\nu$ and $E_\nu$ are the neutrino momentum and energy, 
and $|M_{\rm f\,i}|^2$ is the squared matrix element summed over spin states. 
In superfluid matter the matrix elements acquire coherence
pre-factors (Bogoliubov coefficients) because the quasi-particles 
are a superposition of particles and holes. We will neglect
these prefactors, which is valid when $\Delta \ll \mu$
\cite{Sedrakian:2004qd, Sedrakian:2006mq}.

The matrix element, $|M_{\rm f\,i}|$ depends on the magnitude and relative
orientation of the particle momenta and thus cannot be taken out of the
integral. However, because of the strong degeneracy of nucleons and electrons
in nuclear matter, the main contribution to the integral in Eq.\eqn{eq:emm}
comes from the region near the Fermi surface. Therefore, we can set $|\vec
p|=p_F$ in all smooth functions of energy and momenta. Furthermore, in the
approximation where we treat the nucleons as non-relativistic and ignore the
neutrino momenta as well as the electron and the proton momenta (which are all
small in the region of our interest), the matrix elements turn out to be
independent of the relative orientation of the particle momenta
\cite{yakovlev2001neutrino, friman1979neutrino}. This enables us to take the
matrix element out of the integral. Since we will be interested in calculating
the ratio of the the neutrino emissivity rate with and without
superconductivity or superfluidity, 
the matrix element cancels out at leading order in $\mu_\Delta/p_F$ expansion. Interested readers can find the
expression for the matrix element in Eq.~139 of \cite{yakovlev2001neutrino} and
Eq.~36 of \cite{friman1979neutrino}.

\subsection{Effect of superfluidity}
We will consider phases of nuclear matter with neutron superfluidity,
with or without proton superconductivity.
The superfluidity or superconductivity arises from Cooper pairing due to
the attractive nuclear force between nucleons.
Cooper pairing creates a gap in the energy spectrum of the particles 
near the Fermi surface. For nuclear matter in neutron stars the
temperature is far below the Fermi energy, so only the degrees of freedom
close to the Fermi surface are relevant for transport. Their dispersion
relation $E_i(p_i)$ is
\begin{equation}
(E_i-\mu_i)^2=v_{Fi}^2(p_i-p_{Fi})^2+\Delta_i^2 ,
\end{equation}
where $i=n$ or $p$ indexes the nucleon species, and 
$\Delta_i$ is the gap arising from Cooper pairing. For 
electrons and neutrinos we can use the free dispersion relations
$E_e^2=p_e^2+m_e^2$ and $E_\nu$=$p_\nu$.  

We will write the neutrino emissivity $\eps$ as
\begin{equation}
	\eps=R_{\eps}\,\eps_{0},
\label{eq:Rdef}
\end{equation}
where $\eps_{0}$ is the purely thermal emissivity 
(with no external compression oscillations) for non-superfluid matter,
and $R_\eps$ is the ``modification factor'' that takes into account the
effects of gaps in the fermion spectra
and high amplitude effects such as suprathermal enhancement and gap bridging.
As the gap rises, $R_\eps$ drops below 1 because of Boltzmann suppression \cite{yakovlev1994murca} 
but in the presence of compression oscillations that drive the
system out of beta equilibrium $R_\eps$ can be pushed up to 
very large values. The modification function $R_\eps$ for the modified 
Urca process is 
\begin{equation}
\ba{r@{\;}l}
 R_{\eps} = & \dsp \frac{945\,P_{Fn}^3}{92104\,\pi^{13}}
 \int_0^\infty dx_\nu x_\nu^3 \\
\times & \dsp \prod_{i=1}^{3} \left[
  \int_{-\infty}^{\infty}\!dx_{ni}  W(x_{ni},\De_n/T) \right]\\ 
\times &\dsp \int_{-\infty}^{\infty}\!dx_p W(x_p,\De_p/T) \ A\ \bigl[f(X_{-})+f(X_{+})\bigr]\\[3ex]
\times &\dsp f(x_{n1})f(x_{n2})f(-x_{n3})f(-x_p)  \ ,
\ea
\label{eq:mf_murca}
\end{equation}
where
\begin{eqnarray}
 W(x,z) &\equiv& \dfrac{|x|\Theta(x^2-z^2)}{\sqrt{x^2-z^2}}, \\
\Theta(a) &\equiv& \hbox{1 if $a>0$, or 0 if $a<0$}, \\
A &\equiv& 4 \pi \int \prod_{j=1}^5 d\Omega_j \delta^{3}(\vec{P_{\rm f}}-\vec{P_{\rm i}})\ ,\\
 X_{\pm}&\equiv& x_{n3}+x_p+x_\nu-x_{n1}-x_{n2} \pm \mu_\Delta/T \ ,\\ 
\mu_\Delta &\equiv& \mu_n-\mu_p-\mu_e \ , \label{eq:mudef}\\
f(x) &\equiv& 1/(1+e^{x}) \ ,\\
x_i  &\equiv& (E_i-\mu_i)/T \ .
\end{eqnarray}
The subscripts $n1$ and $n2$ refer to the incoming neutrons;
$n3$ and $p$ refer to the outgoing neutron and proton.
The chemical potential $\mu_\Delta$ arises from external
compression oscillations that drive the system out of $\beta$ equilibrium.
$R_\eps$ is normalized so that $R_\eps=1$ when all the gaps are zero and $\mu_\Delta=0$.

We will calculate $R_{\bar{\eps}}$, which measures how much the emissivity
is affected by nonlinear high amplitude effects and by Cooper pairing,
\begin{equation}
\ba{rcl}
 R_{\bar{\eps}}(\hat{\mu}_\Delta) &=& \dfrac{\< \eps(\mu_\Delta) \>}{\< \eps_0\>}
 = \bigl\langle R_\eps(\mu_\Delta)\bigr\rangle\ , \\[2ex]
\mu_{\Delta}(t) &=& \hat{\mu}_\Delta \sin(\om t)\ ,
\ea
\label{eq:Rbar}
\end{equation}
where $\<X\>$ means the average of $X$ over one oscillation cycle.
Since $\eps_0$, the emissivity in the absence of oscillations and
with no Cooper pairing, is independent of $\mu_\Delta$, then $\<\eps_0\>=\eps_0$.

\section{Singlet state (\osz) pairing for both nucleons}
\label{sec:1S0}

We first consider modified-Urca neutrino emission
in matter where both the neutrons and protons form Cooper pairs
in the \osz\ state. 
As we will see in Sec.~\ref{sec:3P2}, the results for the realistic
case of \tpt\ neutron pairing are qualitatively and quantitatively similar
to this case.
For \osz\ pairing the gap is isotropic so the angular integral 
$A$ and the radial integral in Eq.~\eqn{eq:mf_murca} can be separated. 
After angular integration, the modification function  is

\begin{equation}
\begin{split}
R_{\eps} = & \dsp \frac{60480}{11513\pi^8 }\int_0^\infty dx_\nu x_\nu^3 \\
\times & \dsp \prod_{i=1}^{3}
  \int_{-\infty}^{\infty} dx_{ni} W(x_{ni},\De_n/T) \\
\times & \dsp \int_{-\infty}^{\infty}dx_{p} W(x_p,\De_p/T) \\
\times & f(x_{n1})f(x_{n2})f(-x_{n3})f(-x_p)\bigl[f(X_{-})+f(X_{+})\bigr]\ .
\end{split}
\label{eq:mf_1s0}
\end{equation}

\begin{figure}
\bc
\includegraphics[width=\hsize]{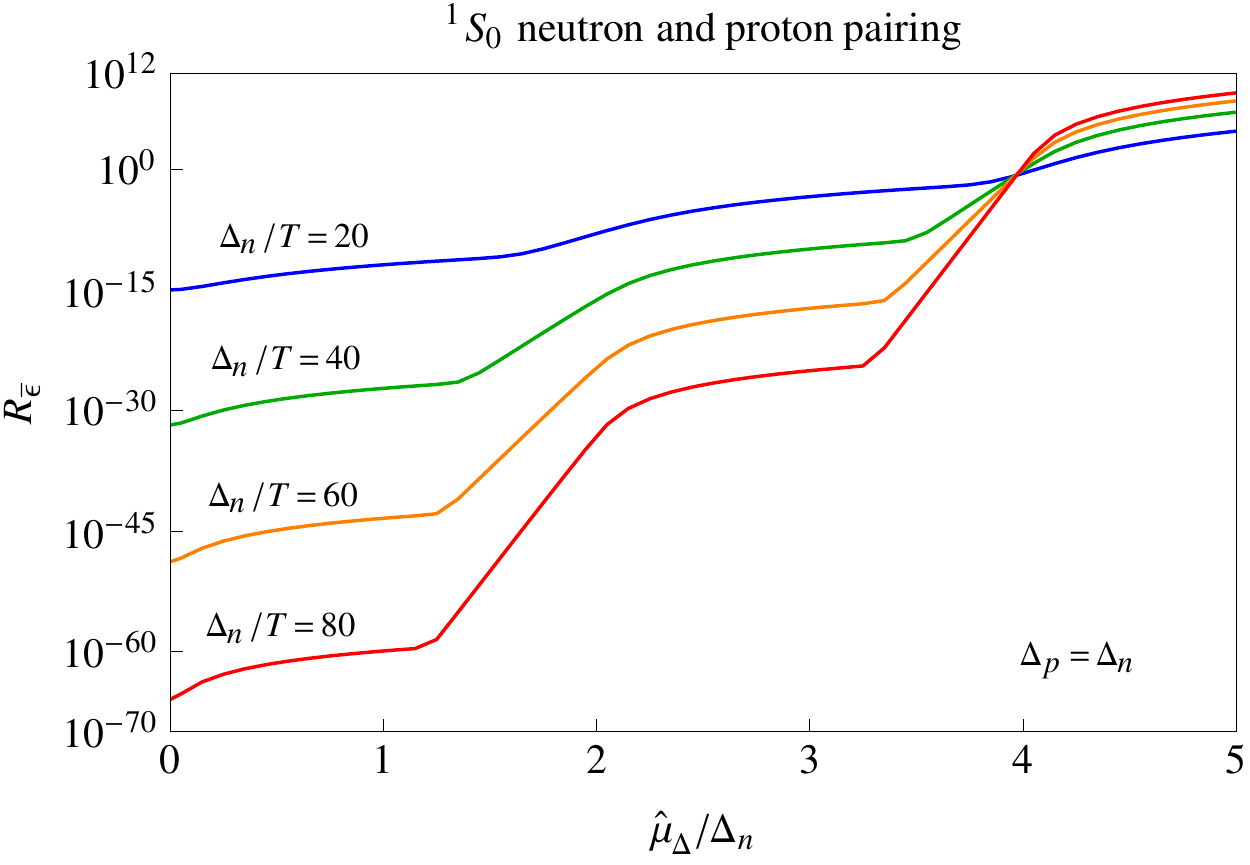}
\caption{
Dependence of neutrino emissivity [via the 
averaged modification
function $R_{\bar \eps}$, Eqs. \eqn{eq:Rbar} and \eqn{eq:Rdef}] on the
amplitude of the applied oscillation [via the departure $\mu_\Delta$ from
$\beta$ equilibrium, Eq. \eqn{eq:mudef}], for \osz\  neutron and proton pairing.
At low amplitude the  emissivity is Boltzmann suppressed by the gaps,
but for high enough amplitude gap bridging reverses the suppression.
}
\label{fig:murca_ngpg}
\ec
\end{figure}

In Fig.~\ref{fig:murca_ngpg} we show the effect of increasing the amplitude
of the applied compression oscillations for the case where protons and
neutrons have the same \osz\  gap, $\De_p=\De_n$.
For low amplitude oscillations the system remains in $\beta$ equilibrium
($\hat{\mu}_\Delta/\De_n \ll 1$) and the neutrino emissivity is very heavily
suppressed by the gaps, roughly as $\exp(-2\De_n/T)$.
As the amplitude rises, $R_{\bar{\eps}}$ rises due to suprathermal effects
[a factor of $(\mu_\Delta/T)^8$ \cite{Reisenegger:1994be}]. 
On the log scale used in 
Fig.~\ref{fig:murca_ngpg} this appears as a very slow, logarithmic, increase.
When $\hat{\mu}_\Delta$ becomes of the same order as $\De_n$, gap bridging begins
to occur: some $\beta$ processes start to become unsuppressed, and their rate rises
exponentially (straight line on this plot). As we discuss in more detail below,
this happens in two steps, until at
high amplitude of the oscillations, $\hat{\mu}_\Delta/\De_n\approx 4$, all the Boltzmann
suppression due to the gap has been overcome, and $R_{\bar{\eps}}\approx 1$,
regardless of how low the temperature may be.

\begin{figure}
\includegraphics[width=\hsize]{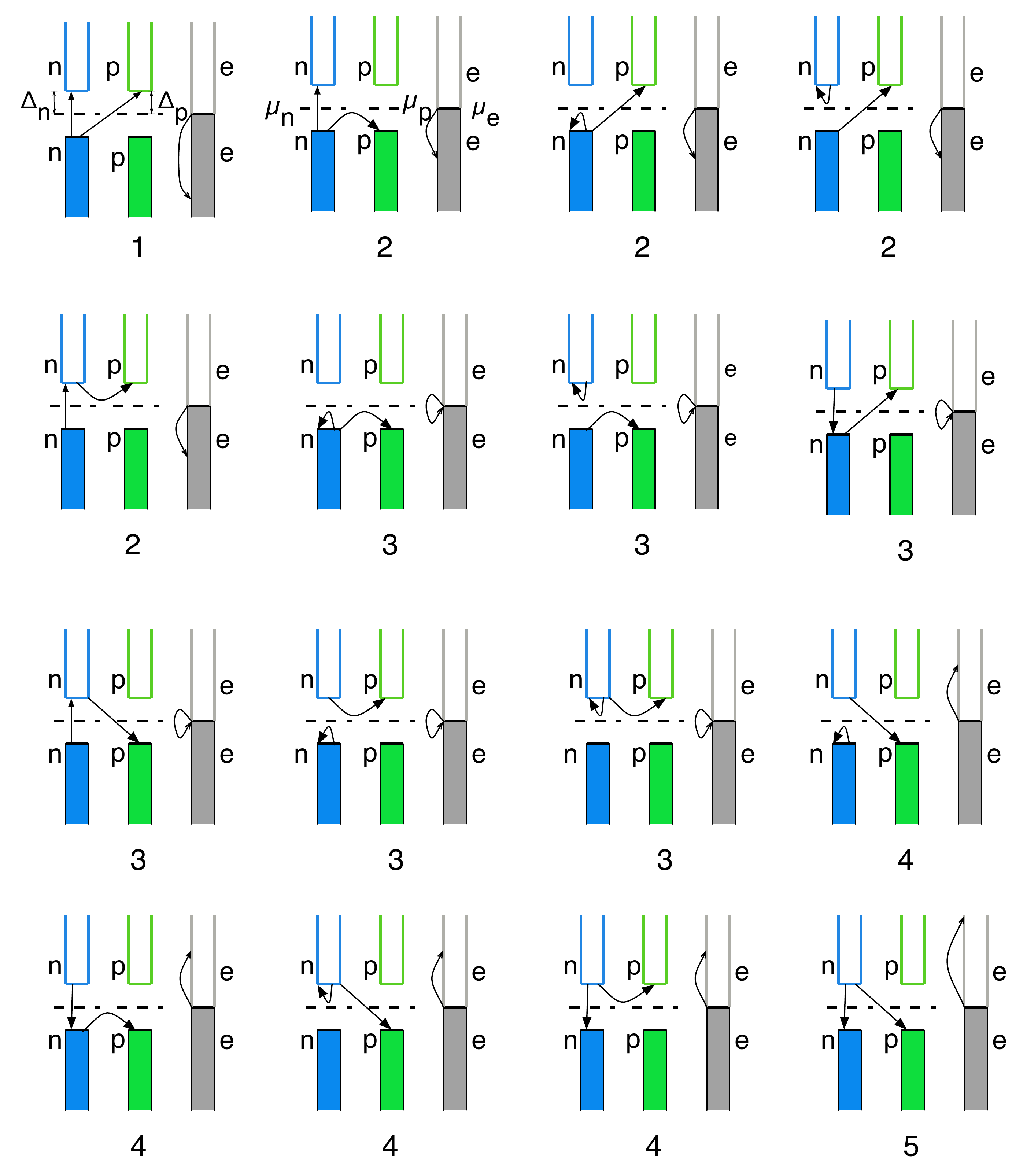}
\caption{The 16 channels that contribute to the modified Urca process.
For each channel we show the three Fermi seas: from left to right,
neutron (gapped), proton (gapped), and electron (ungapped) 
with their Fermi energies aligned.
The black arrows show the transitions of the spectator neutron
(leftmost arrow), the $n \leftrightarrow p$ conversion
(arrow that goes from neutron to proton Fermi sea) and electron
(rightmost arrow). The vertical lengths of the arrows represent the
free energy input or output; by energy conservation these add up to zero
in each process.
}
\label{fig:channels_diag}
\end{figure}

To understand the staircase-like behavior of the dependence of the emissivity
on the amplitude, it is necessary to analyze the different channels that
contribute to the modified Urca process.
These channels are schematically shown in
Fig.~\ref{fig:channels_diag} and their contribution to the modification
function $R_{\bar{\eps}}$ is shown in Fig.~\ref{fig:murca_channels} where we can
already see how gap bridging is manifested at different values of $\hat{\mu}_\Delta$
depending on the channel, so the sum of all the channels yields a staircase
dependence on $\hat{\mu}_\Delta$.

Figure ~\ref{fig:channels_diag} contains 16 panels showing all the channels that
contribute to the modified Urca process. Each of the two nucleons 
can have an initial state above or below its Fermi energy
and a final state above or below its Fermi energy, 
yielding $2^4=16$ possibilities.
For each channel we show the three Fermi seas in $\beta$ equilibrium: 
from left to right, 
neutron (gapped), proton (gapped), electron (ungapped).
We are interested in the free energy $\eps_i-\mu_i$ of each species $i$,
so the Fermi energies are aligned. The black arrows show the
transitions of the spectator neutron, the $n \leftrightarrow p$ conversion
(arrow that goes from neutron to proton Fermi sea), and electron.  The
vertical length of each arrow shows the free energy input or output, so by energy
conservation and beta equilibration ($\mu_n=\mu_p+\mu_e$)
these add up to zero in each process. The electron line starts at the
electron Fermi energy because the free energy cost of placing an 
electron there is zero. The length of the electron arrow then tells us the
amount of energy yielded (or consumed) by the hadronic processes.

\subsection{Rates at zero compression amplitude}

The processes fall into five classes, labeled by the number below each panel.
At $\hat{\mu}_\Delta=0$ the degree of suppression of the rate
for each process can be estimated by keeping track of the
Boltzmann factors that arise when one tries to annihilate a fermion
in a sparsely populated part of phase space, or create a fermion
in a densely occupied part of phase space, according to the following rules:
\beq
\parbox{0.8\hsize}{
$\bm{-}$ For each arrow starting at energy $+|E|$ (i.e.~above the Fermi surface):
a factor of $\exp(-|E|/T)$. \\[1ex]
$\bm{-}$ For each arrow ending at energy $-|E|$ (i.e.~below the Fermi surface):
a factor of $\exp(-|E|/T).$
}
\label{eq:dmu0rule}
\eeq
The result of applying this rule to each class of channel is:

\beq
\ba{r@{\qquad}l}
\mbox{Classes 1 and 5:} & \exp(-4\De/T) \\
\mbox{Classes 2 and 4:} & \exp(-3\De/T) \\
\mbox{Class 3:} & \exp(-2\De/T)
\ea
\eeq
For example, class 1 contains one channel (top left panel of
Fig.~\ref{fig:channels_diag}). To see that this has
a suppression factor of $\exp(-4\De/T)$, we look at each arrow in turn.
The spectator neutron transitions from
an occupied state at free energy $-\De_n$ to an unoccupied state at
free energy $+\De_n$, so it contributes no suppression factor according to
the rules. The ``protagonist'' neutron starts in an occupied
state at free energy $-\De_n$ and becomes a proton in an unoccupied state
at free energy $+\De_p$, so it also contributes no suppression factor.
However, these two processes each require an energy input of $2\De$ so
the electron that is created by the $\beta$ process must produce $4\De$
by ending up in a state at free energy $-4\De$ which is deep in the occupied
electron sea, yielding a suppression factor of $\exp(-4\De/T)$ 
which reflects the unlikeliness of finding an unoccupied state there.

We can now understand the $\hat{\mu}_\Delta=0$ part of Fig.~\ref{fig:murca_channels}:
channels 1 and 5 are the most suppressed, then channels 2 and 4,
and finally the least suppressed is channel 3 where the
nucleon transitions are energy neutral so the electron yields or requires
no energy.

\subsection{Rates as a function of compression amplitude}

\begin{figure}
\bc
\includegraphics[width=\hsize]{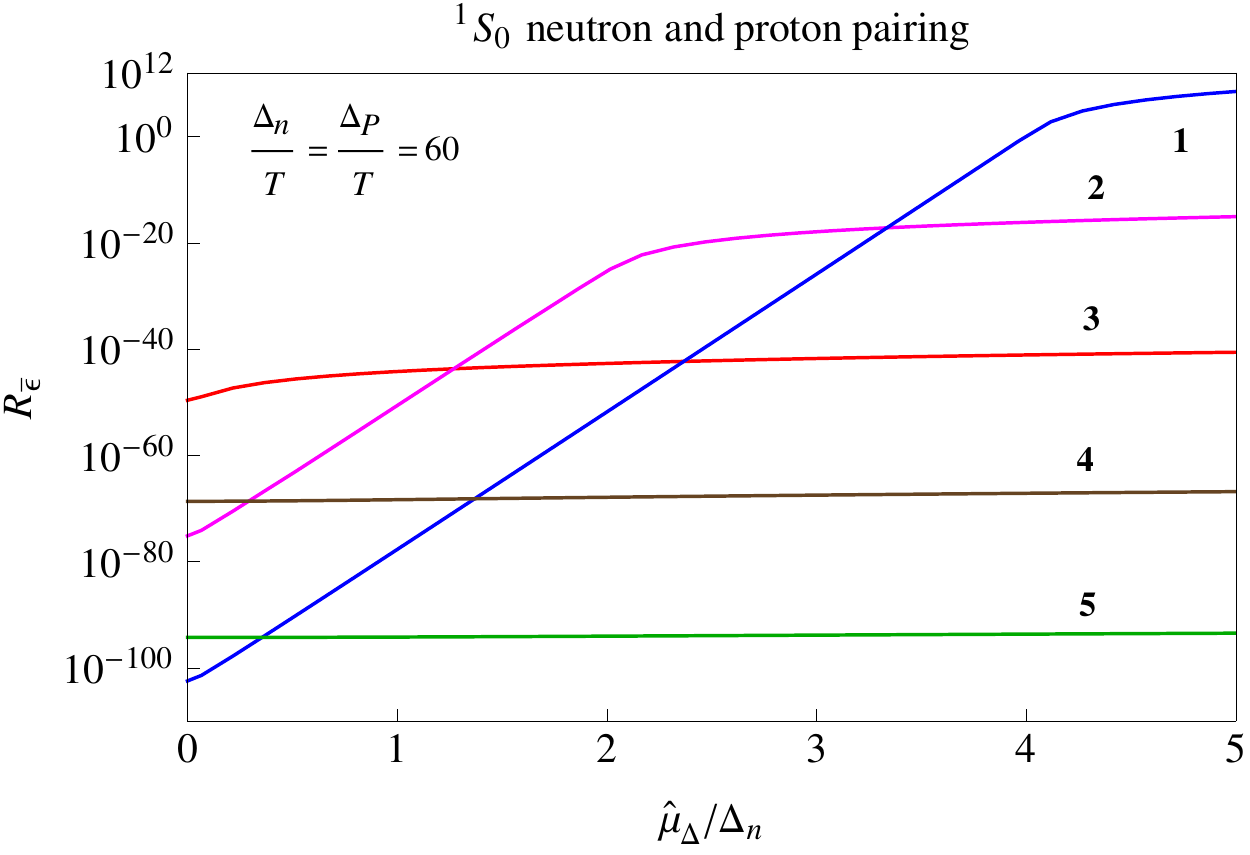}
\caption{Amplitude dependence of the neutrino emissivity for different channels that 
contribute to the modified Urca process. This explains the step structure
in Fig.~\ref{fig:murca_ngpg}}
\label{fig:murca_channels}
\ec
\end{figure}

We now analyze the variation of the rates in the various channels in the
presence of a density oscillation whose amplitude $\hat{\mu}_\Delta$
rises to values as large as $5 \De_n$. The external compression drives
the system out of $\beta$ equilibrium. Under a relatively fast
compression the proton and neutron Fermi energies increase by the same 
fraction, but because the proton fraction in
$\beta$-equilibrated matter rises with density, the proton Fermi energy
is then $\mu_\De$ below its $\beta$-equilibrated value at the higher density.
This reflects the fact that the system now
``wants to make more protons''. 

The rules are modified as follows,
\beq
\parbox{0.85\hsize}{
$\bm{-}$ For channels where the electron is created with energy
$-|E|$ (i.e.~below the electron Fermi energy), its Boltzmann factor is now: \\
$\exp((-|E|+\mu_\Delta)/T)$, or $(\mu_\Delta/T)^8$ if $\mu_\De>E$. \\[1ex]
$\bm{-}$ For channels where the electron is created with energy
above the electron Fermi energy, there is an additional overall
factor of $(\mu_\De/T)^8$ (suprathermal enhancement).
}
\label{eq:dmurule}
\eeq


 In channels of class 3-5, at $\hat{\mu}_\Delta=0$ the electron is created in a state at or above the Fermi energy. There is therefore no gap-bridging, just suprathermal enhancement which multiplies the $\hat{\mu}_\Delta=0$ rate by $(\mu_\De/T)^8$. On the log scale used in Fig. \ref{fig:murca_channels} this gives a very weak growth with $\hat{\mu}_\Delta$ which appears as the horizontal lines at approximately 
$\exp(-2\Delta/T)$ (class 3 channels), $\exp(-3\Delta/T)$ (class 4 channels) and $\exp(-4\Delta/T)$ (class 5 channels).
For class 3 channels, where the electron is created at its
Fermi energy, there is a small gap-bridging growth at very small amplitudes
where $\mu_\De\sim T$ .

In class 2 channels, at $\hat{\mu}_\Delta=0$ the hadronic processes are suppressed
in two ways. Firstly, there is a Boltzmann factor of $\exp(-\De/T)$ from
either trying to place a final-state hadron in the mostly occupied states
below the gap, or from finding an initial-state hadron in the
sparsely occupied states above the gap. Secondly, the hadronic
processes require an
energy input of $2\De$, either to move the spectator neutron up above its
pairing gap, or to convert the other neutron into
a proton above its pairing gap. This leads to an additional suppression
by  $\exp(-2\De/T)$ since to deliver this amount of energy the electron
must be created at free energy $-2\De$, deep in its  occupied Fermi sea.
As $\hat{\mu}_\Delta$ increases, this second factor is canceled by
gap bridging: the required energy input is
reduced by $\hat{\mu}_\Delta$, since the proton Fermi sea is lowered by this amount.
The electron can therefore be created in a 
state with free energy $\hat{\mu}_\Delta-2\Delta$, so the second Boltzmann suppression
factor is reduced and eventually 
when $\hat{\mu}_\Delta \approx 2\Delta$
the electron has enough energy to be placed in a state
above its Fermi energy where there are plenty of unoccupied states and
there is no Boltzmann suppression factor.
Any further increase in $\hat{\mu}_\Delta$ only results in suprathermal enhancement, with
the remaining $\exp(-\De/T)$ (described at the start of this paragraph)
which is not affected by gap bridging.

In channel 1, 
all the hadrons are taken from below the gap and created
above the gap, so there are no Boltzmann factors associated with hadronic
Fermi-Dirac distributions. However, this requires a large energy
input ($4\De$ at $\hat{\mu}_\Delta=0$) from the electron,
which leads to a suppression factor
of $\exp(-4\De/T)$ from forcing the electron into the heavily occupied
phase space deep in its Fermi sea at a free energy of $-4\De$.
As $\hat{\mu}_\Delta$ rises, the proton energy states drop relative to the neutron
ones, and the $n\to p$ process requires less and less energy.
At $\hat{\mu}_\Delta=2\De$ the process breaks even, and at $\hat{\mu}_\Delta=4\De$ it
can provide all the energy needed by the spectator neutron. The electron
is then relieved of the requirement to subsidize the hadrons, and can be
created above its Fermi energy. All Boltzmann suppression has been canceled
by gap bridging, and further increase in $\hat{\mu}_\Delta$ produces only
suprathermal enhancement. 
In conclusion, even though this channel is the most suppressed at $\hat{\mu}_\Delta=0$ it dominates at large $\hat{\mu}_\Delta$ because all the suppression arises from the
hadronic energy requirements, which can be eliminated by gap bridging.

\begin{figure}
\bc
\includegraphics[width=\hsize]{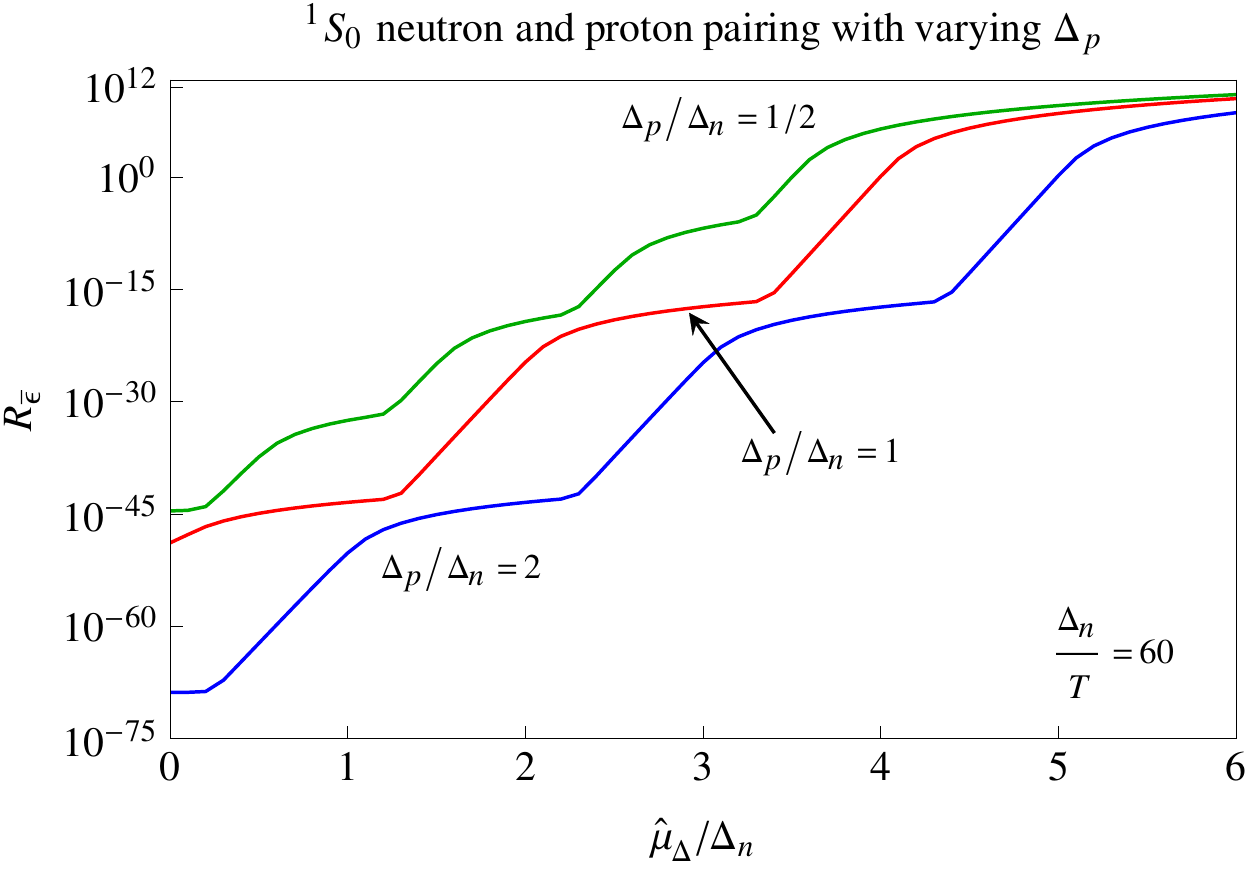}
\caption{How the amplitude dependence of the neutrino emissivity
depends on the \osz\ proton pairing gap, for
\osz\ neutron pairing with $\De_n/T=60$.
We show curves for $\De_p/\De_n=  1/2,\ 1,\ 2$.}
\label{fig:murcagb_n60}
\ec
\end{figure}

Up to now we have assumed that the protons and neutrons have the
same pairing gap.
We now explore the effect of varying the
proton gap: fixing the temperature so that $\De_n/T=60$, we
plot the amplitude dependence of the emissivity for
$\De_p/\De_n = 2,\ 1,\ 1/2,$.

In Fig.~\ref{fig:murcagb_n60} we show the results of this variation in
$\De_p$. Not surprisingly, lower values of the proton gap bring the point of
complete gap bridging to a lower $\hat{\mu}_\Delta/\Delta_n$. This is because at
large $\hat{\mu}_\Delta/\Delta_n$ the rate is dominated by channel 1 where, as
explained in the previous paragraph, all the suppression comes from the energy
requirements of the hadronic processes.  For the spectator neutron to jump the
gap requires $2\De_n$ and the conversion of a neutron below the gap to a
proton above the gap requires energy $\De_n+\De_p$, so the total requirement
is $3\De_n+\De_p$. Thus when $\De_p/\De_n=1/2$ full gap bridging can be achieved when
$\hat{\mu}_\Delta/\De_n\approx3.5$ (top curve in Fig.~\ref{fig:murcagb_n60}) rather than
$4\De_n$ when $\De_p=\De_n$ (middle curve in Fig.~\ref{fig:murcagb_n60})

When $\De_p/\De_n = 2$, the suppression factor is $\exp(-5\Delta_n/T)$ so a compression oscillation of magnitude $\hat{\mu}_\Delta/\Delta_n \approx 5 $, is necessary for complete gap bridging. 


\section{Triplet state (\tpt) neutron pairing}
\label{sec:3P2}

We now calculate the modified Urca neutrino emissivity for nuclear matter in
the inner regions of a neutron star where the neutron density is high and the
neutrons pair in a \tpt\ channel, while the proton density is relatively
low and the protons pair in a ${}^1\!S_{0}$ channel
\cite{takatsuka1971superfluid}.  For the neutrons there are other available
triplet channels, but ${}^3\!P_0$ is only weakly attractive while the
${}^3\!P_1$ state is repulsive \cite{dean2003pairing,tamagaki1970superfluid}.
For the \tpt\ channel, there is still a choice of orientation of the
condensate: $J_z$ could be $0,\,\pm 1,\,\pm 2$. 
Microscopic calculations
\cite{1992NuPhA.536..349B,takatsuka1971superfluid,amundsen1985superfluidity}
find that $J_z=0$ is very slightly energetically favored over the other
values, however this is not conclusive because of
uncertainties in the microscopic theory
\cite{potekhin2015neutron}.
In the following discussion we will consider neutron condensates
with $J_{z}=0$ and $\pm 2$.
We expect these to show different dependencies of the emissivity on
temperature and oscillation amplitude because for $J_z=0$
all neutron states at the Fermi surface are gapped, 
but for $J_{z}=\pm 2$ there are ungapped nodes at the poles
\cite{yakovlev2001neutrino}.

In our calculations we will assume $\De_p=\De_n$ for simplicity.
In real neutron star core matter it is likely that
$\De_p$ is significantly larger than $\De_n$ \cite{Ding2016}.

\subsection{\tpt ($J_z=0$) neutron pairing}


For neutrons that Cooper pair in the \tpt\ state with $J_{z}=0$, 
rotational symmetry is broken. There is a preferred direction in space
(we will align the $z$ axis along it) and the
gap in the neutron spectrum becomes dependent on the angle $\th$ between
the momentum of the neutron quasiparticle and the $z$ axis
\cite{takatsuka1971superfluid},
\begin{equation}
\Delta_n(\th)=\Delta_{n0}\sqrt{1+3\,\cos^2(\theta)} \ .
\end{equation}
Note that the gap varies between a minimum of
$\De_{n0}$ (around the equator) and $2\De_{n0}$ (at the poles)
but does not vanish
anywhere on the Fermi surface. We therefore expect that \tpt\ pairing
will be qualitatively similar to \osz\ pairing, having
the same parametric dependence of the rate on temperature and 
oscillation amplitude.

The dependence of the \tpt\ gap on $\theta$ restricts us from separating the angular and radial part of the integral in Eq. (\ref{eq:mf_murca}). However, the gap has no $\phi$ dependence so we can integrate the momentum-conserving $\delta$ function in Eq. (\ref{eq:mf_murca}) over the azimuthal angles analytically 
 \cite{gusakov2002neutrino}
\begin{equation}
\ba{r@{}l}
\dsp\int_{0}^{2 \pi} & d\phi_1 d\phi_2d\phi_3\delta^3(P_1+P_2+P_3)=\\[2ex]
 & \dsp \frac{4\pi\, \Theta\bigl(\tfrac{3}{4}-c_1 c_2-c_1^2-c_2^2\bigr)}{
 p_{Fn}^3\sqrt{\tfrac{3}{4}-c_1c_2-c_1^2-c_2^2}}\delta(c_1+c_2+c_3)
\ea
\end{equation}
where $c_j\equiv \cos(\theta_j)$ and $\Theta$ is the unit step function. 
We then perform the remaining angular and radial integrals numerically.

\begin{figure}
\bc
\includegraphics[width=\hsize]{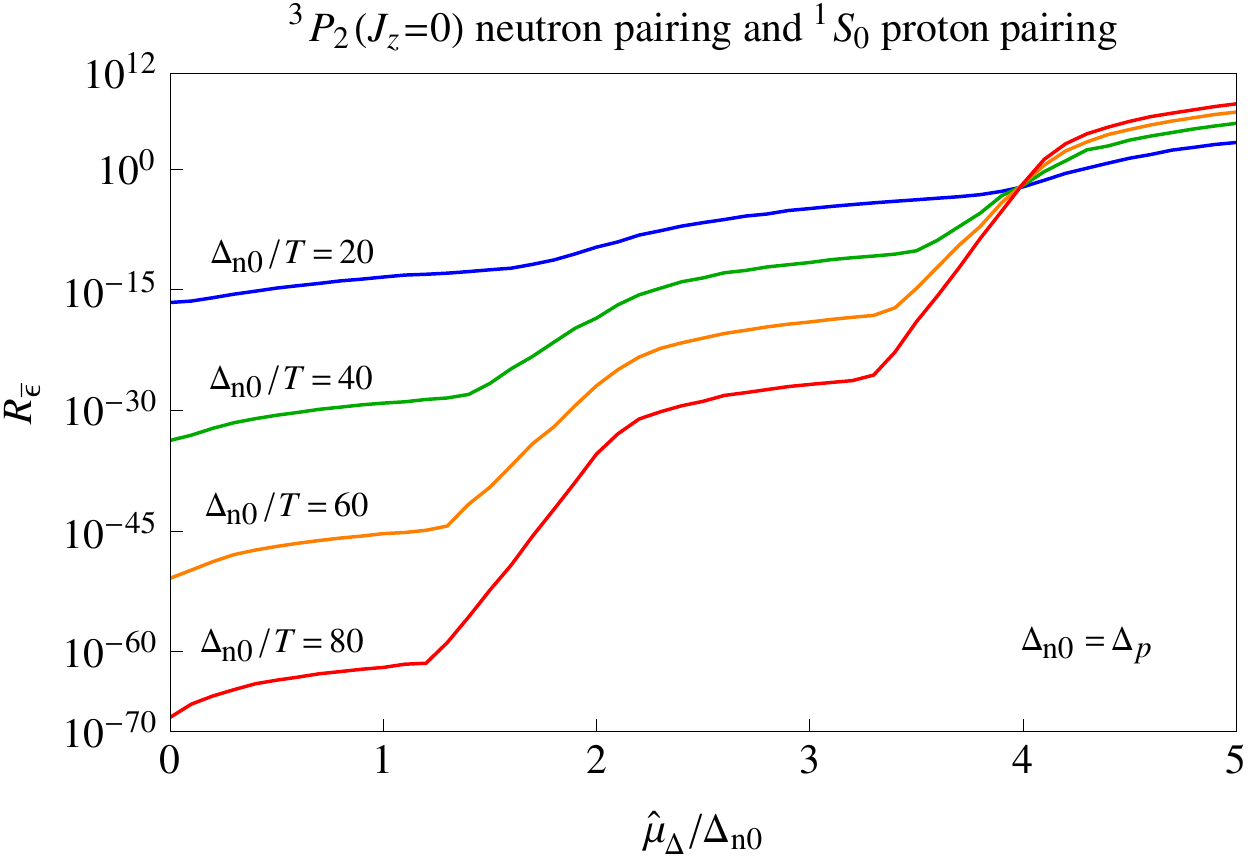}
\caption{
Dependence of the neutrino emissivity on the amplitude of the applied 
compression oscillations, for \tpt\ $(J_z=0)$ neutron pairing and \osz\ proton pairing.
}
\label{fig:murca_m0_gb}
\ec
\end{figure}

Fig.~\ref{fig:murca_m0_gb} shows how the neutrino emissivity is affected
by increasing the amplitude of compression oscillations. 
Since the neutrons are gapped everywhere on the Fermi surface
we expect the results to be similar to those calculated for \osz\
neutron pairing in Sec.~\ref{sec:1S0}, and comparing
Fig.~\ref{fig:murca_m0_gb} with Fig.~\ref{fig:murca_ngpg} we see this
this is indeed the case. The overall pattern of the dependence on
$T$ and $\hat{\mu}_\Delta$ is the same, the only difference is that
$R_{\bar{\eps}}$ for  \tpt$(J_z=0)$ pairing is smaller than for 
\osz\ pairing by a factor that varies between 400 and 20 
as $\hat{\mu}_\Delta$ ranges from 0 to 5$\Delta_n$.


\subsection{\tpt ($J_z=\pm 2$) neutron pairing}

We now consider the case where the neutron Cooper pairs are in 
the \tpt\ state with $|J_z|=2$ while the protons pair in the \osz\ channel. 
The angular dependence of the neutron gap in this channel is \cite{takatsuka1971superfluid}
\begin{equation}
\De_n(\th)=\De_{n0} \sin(\theta)
\end{equation}
Note that the neutron gap vanishes at the poles and has a maximum value of $\De_{n0}$ around the equator.

\begin{figure}
\bc
\includegraphics[width=\hsize]{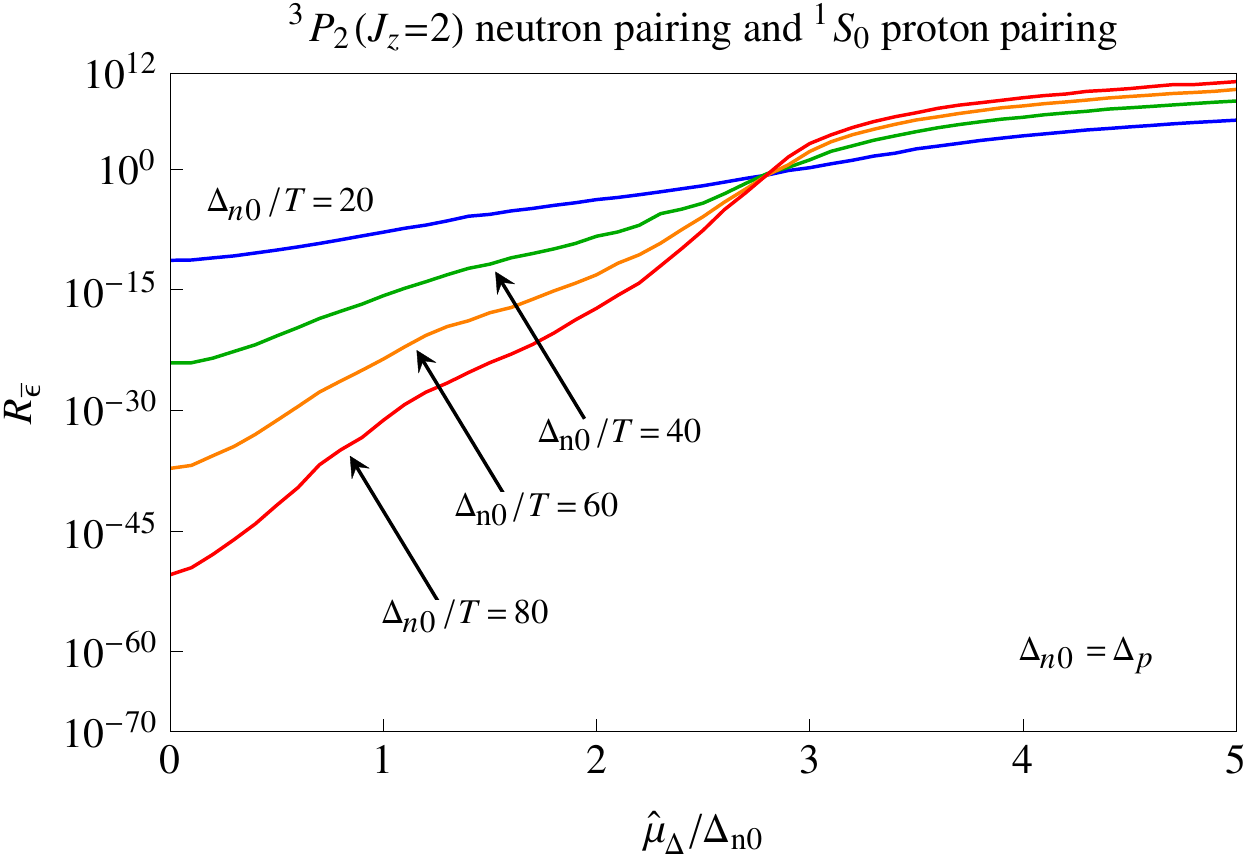}
\caption{
Dependence of the neutrino emissivity on the amplitude of the applied 
compression oscillations, for \tpt\ $(J_z=2)$ neutron pairing 
and \osz\ proton pairing.}
\label{fig:murca_m2_gb}
\ec
\end{figure}

\begin{figure}
\bc
\includegraphics[width=0.7\hsize]{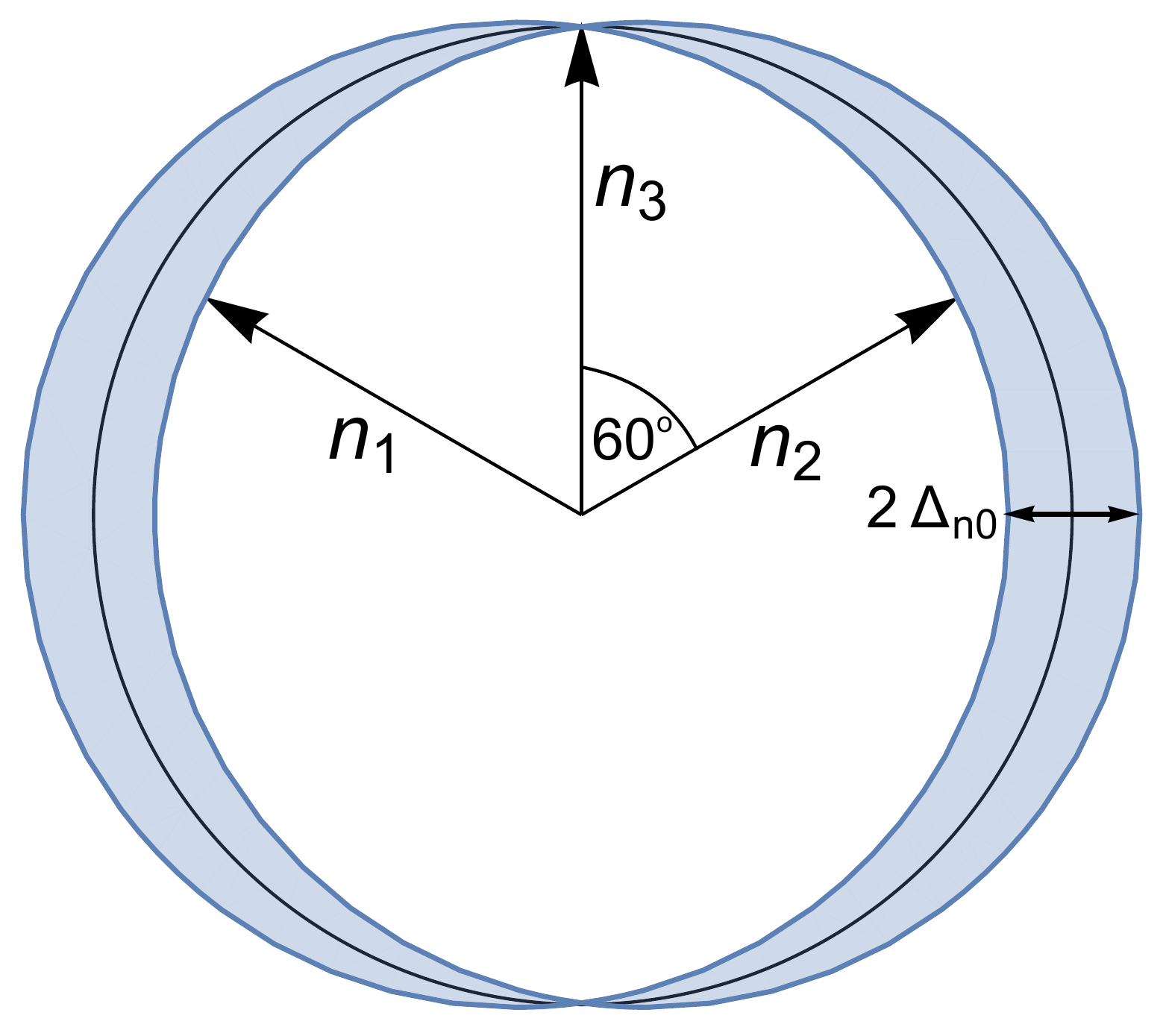}
\ec
\caption{
Momenta of the neutrons in one example of a
typical Urca process in a ${}^3P_s (J_z=\pm 2)$ neutron superfluid.
The shaded region is the gap at the neutron Fermi surface.
The momenta of the two initial state neutrons $n_1$ and $n_2$
add up to the momentum of 
the final state neutron $n_3$ (neglecting the proton and electron momenta). 
Only one of the three neutron momenta can be at a gapless node
on the Fermi surface.
}
\label{fig:Jz2-momenta}
\end{figure}

In Fig.~\ref{fig:murca_m2_gb}, we show the effect of increasing the amplitude
of the applied compression oscillations. To explain this figure
we first explain how the angular dependence of the
gap affects the modified Urca process. In Fig.~\ref{fig:Jz2-momenta}
we have plotted a typical arrangement of the neutron momenta.
To understand the overall behavior we can neglect the proton
and electron Fermi momenta, which are significantly smaller than the
neutron Fermi momentum. The momenta of the incoming neutrons $n_1$ and $n_2$
(Fig.~\ref{fig:Feynman})
then add up to the momentum of the outgoing neutron $n_3$. 
Since all three neutron momenta lie near the
neutron Fermi surface, $p_{n1}$ and $p_{n2}$ must be at a $60^\circ$ angle
to $p_{n3}$, on opposite sides in the same plane. 
Only one of the three neutrons can be at the gapless node on their Fermi 
surface. In Fig.~\ref{fig:Jz2-momenta} we placed $p_{n3}$ at the node. 
The other two neutron momenta
are at $\th=\pi/3$ where the gap is $\sqrt{3}\De_{n0}/2$.

We can now understand 
the effect of increasing the amplitude
of the applied compression oscillations, as shown
in Fig.~\ref{fig:murca_m2_gb}. For low amplitudes the neutrino
emissivity is exponentially suppressed by the gap, roughly as 
$\exp(-1.73 \De_{n0}/T)$, as compared with $\exp(-2\De_{n}/T)$
for \osz\ neutron pairing (Fig.~\ref{fig:murca_ngpg}). 
It is natural to expect the
\tpt$(J_z=\pm 2)$ pairing to be less suppressed because there
is a gapless node on the neutron Fermi surface.

\begin{figure}
\bc
\includegraphics[width=\hsize]{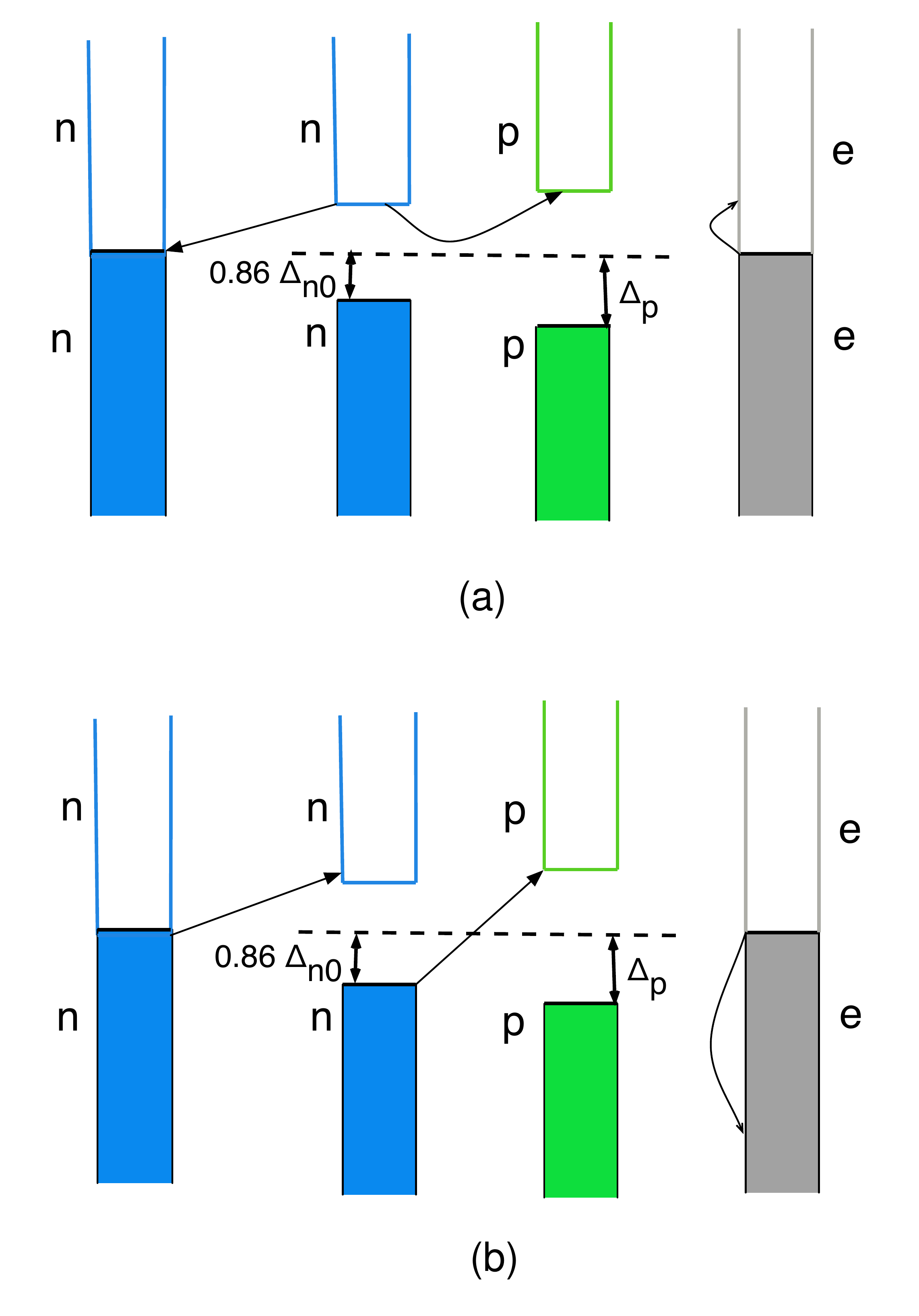}
\caption{
Channels that in the presence of \tpt($J_z=\pm 2$) neutron pairing
and \osz\ proton pairing,  will dominate the modified Urca process 
in different regimes:
(a) at low $\mu_\Delta/\De_{n0}$, (b) at high $\mu_\Delta/\De_{n0}$.
}
\label{fig:murca_m2_channel}
\ec
\end{figure}

To understand more fully the origin of the suppression factor, we analyze 
one of the dominant channels at low $\hat{\mu}_\Delta/\Delta_{n0}$,
shown in Fig.~\ref{fig:murca_m2_channel} (a). 
In the figure we show two neutron Fermi seas, one gapless,
corresponding to a neutron momentum at the gapless node on the Fermi
surface (polar angle $\th=0$), 
and one with a gap of $\sqrt{3}\De_{n0}/2\approx 0.866 \De_{n0}$, 
corresponding to
neutron momenta at $\th=\pi/3$ or $2\pi/3$.
For the process shown in Fig.~\ref{fig:murca_m2_channel} (a) there is
a suppression factor of $\exp(-\sqrt{3}\De_{n0}/t)\approx\exp(-1.73 \De_{n0}/t)$, arising from the unlikelihood of finding both initial
state neutrons in the sparsely occupied region above the gap at
$\th=\pi/3$. The final state neutron is at the gapless node, and so 
experiences no Boltzmann suppression. The electron is created above its
Fermi surface (because the hadrons provide an energy surplus to be
absorbed by the electron) so it too experiences no Boltzmann suppression.

As we see in Fig.~\ref{fig:murca_m2_gb}, increasing the
amplitude of compression oscillation leads to gap bridging:
some of the $\beta$ processes 
become unsuppressed which results in steady increase of
the neutrino emissivity until 
at $\hat{\mu}_\Delta/\De_{n0} \approx 2.73$
the $\beta$ process rate reaches the ungapped limit $(R_\eps\sim 1)$, 
regardless of how low the temperature may be. 

To understand why complete gap bridging happens at
$\hat{\mu}_\Delta/\De_{n0} \approx 2.73$, we show in
Fig.~\ref{fig:murca_m2_channel} (b) the channel of class 1
(Fig.~\ref{fig:channels_diag}) which
dominates at large $\hat{\mu}_\Delta/\Delta_{n0}$.
As in the case of \osz\ neutron pairing, this is
because all the suppression comes
from the energy requirements of the hadronic sector, none from
hadronic Fermi-Dirac factors (all hadrons start below the gap
and end in the sparsely occupied region above the gap). 
At $\hat{\mu}_\Delta=0$ the energy
required for this is $0.866\De_{n0}$ for $n_1\to n_3$ and
$0.866\De_{n0}+\De_{p}$ for $n_2\to p$, totalling
$2.73\De_{n0}$ assuming $\De_p=\De_{n0}$. This energy is supplied
by the electron, which must be created deep in its Fermi sea at
a free energy of $-2.73\De_{n0}$, yielding a Boltzmann suppression
of $\exp(-2.73\De_{n0}/T)$.

As we increase $\hat{\mu}_\Delta$, the proton Fermi surface is lowered by
$\hat{\mu}_\Delta$ relative to the neutron Fermi surface
and the $n\to p$ process requires less and less energy.
At $\hat{\mu}_\Delta=1.866\De_{n0}$ the $n\to p$
process breaks even, and at $\hat{\mu}_\Delta=2.73\De_{n0}$ it
can provide all the energy needed by the spectator neutron. The electron
can then be created above its Fermi energy. 
All Boltzmann suppression has then been canceled
by gap bridging, and further increase in $\mu_\Delta$ produces only
suprathermal enhancement.

\section{Conclusion and discussion}
\label{sec:conclusion}

We have shown that the exponential suppression of flavor-changing
$\beta$ processes in superfluid and superconducting nuclear matter
can be completely overcome, via the mechanism of gap bridging,
by compression oscillations
of sufficiently high amplitude, regardless of how low the temperature
may be. This confirms the conjecture outlined in previous work
\cite{Alford:2011df}, and shows that it applies to the realistic case
of modified Urca processes and \tpt\ neutron pairing.

We expect gap bridging to be relevant in processes that induce density
oscillations of amplitude $\De n/\bar n \sim 10^{-3}$ to $10^{-2}$
\cite{Alford:2011df}. This is
sufficient to overcome typical nucleon pairing gaps which are of order
1\,MeV. In hyperonic \cite{Prakash:1992ApJ...390L..77P,Haensel:2001em} or
quark \cite{Schmitt:2005wg,Wang:2010ydb} matter, there are processes which are
only suppressed by $\Delta\lesssim 0.01$\,MeV, and these could be bridged by
oscillations with amplitudes as small as $\Delta n/\bar{n}\lesssim 10^{-4}$. 
Relevant physical scenarios that are likely
to involve high-amplitude oscillations include
unstable oscillations of rotating compact stars such as $f$-modes or $r$-modes \cite{Stergioulas:2003yp},
events like star quakes \cite{franco2000quaking}, and 
neutron star mergers \cite{tsang2012shattering}.
When such
high-amplitude compression oscillations occur
in superfluid or superconducting matter,
certain transport properties, such as bulk
viscosity and neutrino emissivity, will be greatly enhanced, leading to
nonlinear (in amplitude) damping of the mode itself, and enhanced
cooling via neutrino emission.

\begin{acknowledgments}
We thank Kai Schwenzer for valuable comments. This material is based upon work supported by the U.S. Department of Energy, Office of Science, Office of Nuclear Physics under Award No.
\#DE-FG02-05ER41375. 
\end{acknowledgments}

\bibliography{gapbridging_murca}{}
\bibliographystyle{apsrev4-1}
\end{document}